\documentclass[10pt]{article}
\usepackage{amsmath}
\usepackage{amssymb}

\usepackage{longtable}

\usepackage{graphicx}

\usepackage{color} 

\usepackage[labelfont=bf,labelsep=period,justification=raggedright]{caption}

\usepackage[sorting=none]{biblatex}

\bibliography{ss2,ia}

\makeatletter
\renewcommand{\@biblabel}[1]{\quad#1.}
\makeatother

\date{}

\begin{document}

\begin{flushleft}
{\Large
\textbf{From Social Simulation to Integrative System Design\\}
}
Dirk Helbing$^{1,2}$ and Stefano Balietti$^{1}$ 
\\
\bf{1} ETH Zurich, CLU, Clausiusstr. 50, 8092 Zurich, Switzerland
\\
\bf{2} Santa Fe Institute, 1399 Hyde Park Road, Santa Fe, NM 87501, USA
\\
$\ast$ E-mail: 
{\normalfont dhelbing@ethz.ch and sbalietti@ethz.ch}
\end{flushleft}

\section*{Abstract}

The purpose of this White Paper of the EU Support Action
``Visioneer''\cite{Visioneer_site} is to address the following goals:

\begin{enumerate}
\item Develop strategies to build up social simulation capacities.
\item Suggest ways to build up an ``artificial societies'' community
  that aims at simulating real and alternative societies by means of
  supercomputers, grid or cloud computing.
\item Derive proposals to establish centers for integrative systems
  design.
\end{enumerate}

\section{Introduction}

\subsection{Real-World Challenges}\label{practical}

Since decades, if not since hundreds of years, the world has been facing a number of recurrent socio-economic problems which are obviously hard to solve because their roots are still not well understood. This includes, in particular, the following problems: 
\begin{enumerate}

\item {\it Financial and economic (in)stability} (government debts, taxation, and inflation/deflation; sustainability of social benefit systems; consumption and investment behavior\ldots)

\item {\it Public health} (food safety; spreading of epidemics [flu, SARS, H1N1, HIV], obesity, smoking, or unhealthy diets\ldots)

\item {\it Sustainable use of resources and environment} (consumption habits, travel behavior, sustainable and efficient use of energy and other resources, participation in recycling efforts, environmental protection\ldots) 

\item {\it Collective social behavior} and opinion dynamics (abrupt changes in consumer behavior; social contagion, extremism, hooliganism, changing values; breakdown of cooperation, trust, compliance, solidarity\ldots)

\item {\it Security and peace} (organized crime, terrorism, social unrest, independence movements, conflict, war\ldots)

\item {\it Reliability of critical infrastructures, in particular of information systems} (cyber risks, misuse of sensitive data, espionage, violation of privacy; data deluge, spam; education and inheritance of culture\ldots) 

\item {\it Institutional design} (intellectual property rights; over-regulation; corruption; balance between global and local, central and decentral control\ldots)

\item {\it Balance of power} in a multi-polar world (between different countries and economic centers; also between individual and collective rights, political and company power; avoidance of monopolies; formation of coalitions; protection of pluralism, individual freedoms, minorities\ldots)

\item {\it Demographic change} of the population structure (change of birth rate, migration, integration\ldots)

\item {\it Social, economic and political participation and inclusion} (of people of different gender, age, health, education, income, religion, culture, language, preferences; reduction of unemployment\ldots)

\end{enumerate}
It must be underlined that the above list is not ranked according to importance and that at least some of these challenges are obviously interdependent. A recent paper \cite{Econophysics_Balietti} identifies and discusses a number of fundamental scientific challenges in economics and other social sciences which must be addressed before the above problems can be cured or mitigated (see also Ref. \cite{VisioneerGrandChallenges}). However, there is actually hope that progress can be made, and the following activities are likely to be useful for this:

\begin{enumerate}
\item Build a Financial Crisis Observatory and large-scale agent-based simulations, explore alternative exchange systems (see Ref. \cite{VisioneerInnovationAccelerator}).

\item Create a network of ``health sensors'' and a Health Risk Observatory in order to follow, simulate and anticipate the spreading of diseases; create contingency plans (see Ref. \cite{VisioneerCrisis}).

\item Develop user-friendly ICT tools for a realistic, but pluralistic modeling and simulation of the dynamics of complex systems and create new incentive schemes (see Sec. \ref{WoM}).

\item Explore group dynamics through lab and Web experiments and social data mining (see Ref. \cite{VisioneerInnovationAccelerator}).

\item Understand the roots of conflicts to derive new ways to avoid or mitigate them (see Ref. \cite{VisioneerCrisis}).

\item Employ network analysis, build a privacy-respecting reputation system to learn how to design a self-controlled, trustable system (see \cite{VisioneerCrisis}).

\item Perform evolutionary optimization and scenario analyses, use serious games (see Secs. \ref{shift1} and \ref{shift2}).

\item Develop better coordination and synchronisation mechanisms and new paradigms to manage complexity (see Sec. \ref{managing_complexity}).

\item Create supporting ICT systems to support people in various real-life situations and their mutual interactions (e.g. create a ``multi-national translator or adapter''); use Decision Arenas to identify good solutions (see Sec. \ref{darenas}). 

\item Use eGovernance and build a Knowledge Accelerator (see Ref. \cite{VisioneerInnovationAccelerator}).

\end{enumerate}

Some of the above challenges obviously require massive data-mining activities or the use and development of novel ICT systems, as it has been discussed in other Visioneer White Papers \cite{VisioneerCrisis,VisioneerInnovationAccelerator,VisioneerGrandChallenges}. Of course, we do not want to claim that all problems will be solvable forever, but there are good reasons to believe that substantial progress can be made for some socio-economic issues, and that all possible attempt should be made to mitigate them. The before-mentioned documents explain possible approaches to make such progress. Complementary to these, social super-computing will be needed to make substantial progress in the understanding, management and improvement of complex systems, as outlined in this document.  

\subsection{Particular scientific challenges of socio-economic systems}

We are fully aware of a number of particular challenges of socio-economic systems which make them different from other systems \cite{PluralisticModeling}:
\begin{itemize}
\item the number of variables involved is typically (much) larger, 
\item the relevant variables and parameters are often unknown and hard to measure (the existence of ``unknown unknowns'' is typical),
\item the time scales on which the variables evolve are often not well separated from each other, 
\item the statistical variation of measurements is considerable and masks laws of social behavior, where they exist (if they exist at all),
\item frequently there is no ensemble of equivalent systems, but just one realization (one human history),
\item empirical studies are limited by technical, financial, and ethical issues,
\item it is difficult or impossible to subdivide the system into simple, non-interacting subsystems that can be separately studied,
\item details of the environment and the history of the system, in which the actors live, may be important to understand their behaviour,
\item the observer participates in the system and modifies social reality,
\item the non-linear and/or network dependence of many variables leads to complex dynamics and structures, and sometimes paradoxical effects, 
\item interaction effects are often strong, and emergent phenomena are ubiquitous (hence, not understandable by the measurement and quantification of the individual system elements), 
\item factors such as a large degree of randomness and heterogeneity,  memory, anticipation (future expectations), decision-making, communication, consciousness, and the relevance of intentions and individual interpretations complicate the analysis and modeling a lot,
\item the same applies to human features such as emotions, creativity, and innovation,
\item the impact of information is often more decisive for the behavior of a socio-economic system than physical aspects (energy, matter) or our biological heritage,
\item information feedbacks have a significant impact on the resulting socio-economic system dynamics,
\item the ``rules of the game'' and the interactions in a social or economic system may change over time, in contrast to what we believe to be true for the fundamental laws and forces of physics,
\item in particular, social systems are influenced by normative and moral issues, which are variable.
\end{itemize}

For these reasons, progress in the socio-economic sciences is more difficult than in the natural and engineering sciences. However, many available scientific methods have not yet been used in the socio-economic sciences to the extent they may contribute. This includes, for example, insights from

\begin{itemize}
\item psychology,
\item the neurosciences,
\item biology (particulary the organization of social species and swarms),
\item the ecology of mutually networked entities,
\item synergetics and complex systems theory for large adaptive systems with non-linear interactions (including theories of self-organization phenomena, non-linear and chaotic dynamics, emergence etc.), 
\item statistical physics of systems with many interacting components (including theories of phase transitions and critical phenomena),
\item information theory, 
\item computer science (including data mining, pattern recognition, machine learning,
computer simulation, evolutionary optimization, heuristic algorithms, etc.).
\item artificial intelligence and robotics,  
\item cybernetics/control theory, and
\item game theory.
\end{itemize}

As the methods of these different sciences have by far not been fully explored and utilized, we are convinced that a much better understanding of socio-economic systems can be gained than what we know today. Restricting to empirical, experimental and analytical methods, but underusing the possibilities of modern and future information and communication technologies (ICT) would ignore a large number of powerful research tools. When properly performed, computational approaches can gain results for a much wider range of models, parameters, and conditions than empirical, experimental and analytical methods can. 
 
\section{Strategies to build up social simulation capacities: From agent-based to multi-scale simulations}

According to Shannon \cite{Shannon},
simulation can be defined as ``\textit{the process of designing a
  model of a real system and conducting experiments with this model
  for the purpose of understanding the behavior of the system and/or
  evaluating various strategies for the operation of the system}''. 
The method of computer simulation has been successfully used in the natural and engineering
sciences. Also in the social sciences, it is being used since the 50ies \cite{Orcutt}, 
and it is spreading continuously, particularly since the advent of agent-based modeling (ABM) (for an introduction to history of social science simulation see Chap. 1 of Ref. \cite{GilbertSim4SS}). 

\subsection{Agent-based models and simulation}

Agent-based simulations are used to model the decision-making entities of the system of interest. Depending on the field of application, agents can represent individuals, groups, organizations, companies, etc. They seem to be particularly suited to reflect interactions with the surrounding environment and between different agents as well as their heterogeneous properties in a natural way \cite{Weinberg2002,Wolfram2002}. Agent-based simulations can be combined with a variety of concepts, ranging from decision models to evolutionary game theory and beyond. When modeling individuals, it seems plausible to give agents the following features (or a subset of them, depending on the purpose of the respective study):

\begin{itemize}

\item birth, death, and reproduction,

\item individual needs of resources (e.g. eat and drink),

\item competition and fighting ability,

\item toolmaking ability (e.g. the possibility to grow food, hunt etc.),

\item perception,

\item curiosity, exploration behavior, ability for innovation,

\item emotions,

\item memory and future expectations,

\item mobility and carrying capacity,

\item communication,

\item learning and teaching ability,

\item possibility of trading and exchange,

\item the tendency to have relationships with other agents (e.g. family or friendship ties etc.).

\end{itemize}
 
Agent-based simulations allow one to start off with the descriptive power of verbal argumentation and to calculate the implications of different hypotheses. From this perspective, computer simulation is a special form of mathematics. Some are already arguing that applied computer science has taken the role that mathematics has had for centuries for what concerns  ``providing an orderly formal framework and explanatory apparatus for the other sciences'' \cite{ScienceTwoWays}. Others have formulated reasons why simulation should be considered more appropriate than mathematics for performing social science analysis \cite{TaberTimpone}: modularity, greater flexibility, enhanced expressiveness, and inherent capabilities for performing parallel and distributed tasks are among its strong advantages. 

In the socio-economic sciences, multi-agent computer simulations facilitate to overcome the limitations of the theoretical concept of {\it homo economicus} (the ``perfect egoist'') \cite{Aaron}, relaxing idealized assumptions that are empirically not well supported. They also offer a possibility to go beyond the representative agent models of macroeconomics \cite{KirmanRepresentativeAgent}, and to establish a natural link between the micro- and macro-level of description, considering heterogeneity, spatio-temporal variability, and fluctuations, which are known to change the dynamics and outcome of the system sometimes dramatically \cite{Econophysics_Balietti}. 

Computer simulations are also a suitable tool to study complex systems and their emergent, often counter-intuitive behaviors, due to which linear, conventional, and straight-forward modeling attempts may not provide a suitable understanding \cite{citeulike:3975610,HelbingSystemicRisks_SE}. 
By modeling the relationships on the level of individuals in a rule-based way rather than describing aggregates in the classical, equation-based way \cite{parunak1997go}, agent-based simulations allow peculiar features of a system to emerge without a-priori assumptions on global properties. At the same time, this approach allows one to show the consequences of ex-ante hypotheses regarding the interactions of agents. It is also possible to apply methods from statistics and econometrics to simulation outcomes and compare the results with actual data, to which the same procedures have been applied. As agent-based simulations are suited for detailed hypothesis-testing, one could say that they can serve as a sort of magnifying glass or telescope (``socioscope''), which may be used to understand reality better.

The progress as compared to fifteen years ago lies not only in the better computer power and visualization techniques. The greater perspectives result from the current transition from a data-poor to a data-rich situation in the socio-economic sciences, which allows one to verify or falsify models, calibrate their parameters, or pursue data-driven modeling approaches  \cite{VisioneerCrisis}. This transition supports the comprehension of the inner relationships and significant patterns of complex systems with semi-automated techniques, based on algorithms and ICT systems that are able to cope with the gigantic amounts of data that are now continuously produced by all sorts of information systems (and even more so in the future). Given this development, we envision a new, complementary way of performing socio-economic research, which may be called ``Social Supercomputing''. This approach would facilitate the integration of different kinds of data (e.g. demographic, socio-economic, and geographic data) and different kinds of simulation approaches (e.g. agent-based and continuous ones, using partial differential equations), at an unprecedented scale and level of detail. It would facilitate {\it multi-scale simulations} of all kinds. In fact, global-scale computer simulation is a vision that appears to become possible in the socio-economic sciences within the next 10 to 15 years, with unprecedented opportunities for societies and economies, if done in the right way. 

\subsection{Social super-computing}\label{SSC}

In the past, large or even global-scale simulations of human society have been obstructed by a lack of sufficient computer power and also by a lack of data  to determine suitable models, network interactions, parameters, initial and boundary conditions. However, future high performance computing capacities (including grid and cloud computing) and advances in collecting and analyzing massive socio-economic data \cite{VisioneerCrisis} promise to overcome these problems. Massive computer power is, for example, required for large-scale computational analyses in the following areas:

\begin{itemize}
\item network research, community detection,
\item Monte-Carlo simulations of probabilistic system behavior,
\item multi-agent simulations of large systems (e.g. ``whole-earth
  simulator'', which may involve up to 10 billion agents and
  complementary environmental simulations),
\item multi-agent simulations considering human cognitive and
  psychological processes (e.g. personality, memory, strategic
  decision-making, emotions, creativity etc.),
\item massive data mining, e.g. real-time financial data analysis,
\item realistic computer simulations with parameter-rich models
  (coupling simulations of climate and environmental change with
  simulations of large techno-socio-economic-environmental systems),
\item pluralistic or ``possibilistic'', ``multiple-world-views'' modeling \cite{PluralisticModeling} (to determine
  the degree of reliability of model assumptions and to improve the
  overall prediction capability),
\item calibration of parameter-rich models with massive datasets,
\item scanning of multi-dimensional parameter spaces,
\item sensitivity analyses (e.g. $k$-failures),
\item parallel worlds scenario analyses (to test alternative policies
  etc.),
\item visualization of multi-dimensional data and models of complex
  systems, and
\item an optimal real-time management of complex systems (guided
  self-organization, self-optimization),
\item multi-agent simulations considering complex decision-making processes.

\end{itemize} 
\par
In the past, supercomputers have been used mainly in physics or
biology, or for difficult engineering problems such as the
construction of new aircrafts. They have also been applied to weather
forecasting and the study of climate change
(e.g. \cite{EarthSimulator}). Recently, they are also increasingly used for social and economic analyses, even of the most fundamental human processes. First applications in a variety of fields demonstrate the feasibility of social super-computing and the possibility to apply it in order to better cope with crises:

\begin{enumerate}

\item A first successful application area was large-scale traffic simulation. The TRANSIMS project \cite{citeulike:3578186,Nagel_TRANSIMS_2001,Nagel99transimsfor} of the Los Alamos National Institute (LANL), for example, has created agent-based simulations of whole cities such as Dallas \cite{Nagel97usingmicrosimulation} and Portland \cite{Nagel_Portland}.
The approach has been recently extended to the simulation of the travel behavior of the 7.5 million inhabitants of Switzerland \cite{Axhausen_SwissTravel,Auxausen_frmwk_2005}. These simulations are obviously based on parallel computing. They generate realistic individual activity patterns according to detailed statistical panel data (``travel diaries'') \cite{citeulike:146648,axhausen1992activity} which are nowadays complemented by the availability of GPS data. Other extensions look at the interconnections between the traffic system and the regional development \cite{nagel04,Kai_2010_traffic_Regio}.

\item Recent applications are studying contingency plans for large-scale evacuations of cities \cite{citeulike:5368354,nhess-9-1509-2009}. Key aspect here is understanding the interdependency of infrastructure systems \cite{Towards_Mod_Infr,springerlink:10.1007/3-540-34874-3_14} and their vulnerabilities through natural disasters, terrorist attacks, and accidents. Los Alamos National Laboratories have already established a Critical Infrastructure Protection Decision Support System \cite{NISAC}. Its advanced simulation capabilities have already been extensively used during past emergencies.

\item Large-scale simulations have also been applied to study and predict the spreading of diseases. While previous epidemic spreading models such as the SIR model \cite{citeulike:3766315,citeulike:3766328,citeulike:3766364}
have neglected spatial interaction effects, recent models
\cite{Colizza14022006} take into account effects of spatial proximity \cite{citeulike:5373956}, air traffic \cite{Hufnagel19102004} 
and land transport \cite{citeulike:2180593}
using TRANSIMS and other traffic simulation models. The current scientific
development also tries to take into account behavioral changes which may reduce
the spreading rate of diseases.
\item There are also attempts to model financial markets with agent-based simulations.
Two examples for this are the Santa-Fe Market Simulator \cite{ASM} and U-mart \cite{U-Mart}.
Recent attempts are even heading towards the simulation of the whole economic system (see for example the
EU project EURACE \cite{EURACE}). Other simulation studies are trying to understand the evolution of
cooperation \cite{citeulike:4094207},
social norms \cite{HelbingJohanssonNorms,10.1371/journal.pcbi.1000758},
language \cite{cangelosi2002simulating,dunbar1998grooming,nowak1999evolution},
and culture \cite{epstein1996growing}. Such simulations explore the conditions under which trust, cooperation and other forms of ``social capital'' can thrive in societies (see also Ref. \cite{VisioneerInnovationAccelerator,Econophysics_Balietti}). They also show that the crust of civilization is disturbingly vulnerable. These simulations reveal common patterns behind breakdowns of social order in events as diverse as the war in former Yugoslavia, lootings after earthquakes or other natural disasters, or the violent demonstrations we have recently seen in some European countries.

\item Migration processes have been studied through large-scale
  social simulations as well \cite{citeulike:4301623}. The related scenarios integrate huge volumes of demographic
  information about geographical areas in order to determine the factors that influence
  the spatial population dynamics.
\end{enumerate}

It appears logical that supercomputing will be ultimately moving on from applications in the natural and engineering sciences to the simulation of social and economic systems, as more and more complex systems become understandable and the required data become available. It is obvious that virtual three-dimensional worlds are waiting to be filled with life. The upcoming trend becomes visible already. For example, ETH Zurich's Competence Center for ``Coping with Crises in Complex Socio-Economic Systems'' (CCSS) is currently the biggest shareholder of the Brutus supercomputing cluster, which is ranked 10th on the list of Europe's fastest supercomputers. Social supercomputing is also a new focus of other renowned research centres such as the Los Alamos National Laboratory, the Brookings Institution, John Hopkins University or the Oak Ridge National Laboratory in the United States, to mention only a few. Finally, there is very recent news that Spain is going to invest 2.7 million EUR in a project (Simulpast) aimed at developing new ways of applying simulation and High Performance Computing (HPC) to the Humanities and Social Sciences.

\section{Ways to build up an ``artificial societies'' community
  that aims at simulating real and alternative societies by means of
  supercomputers, grid or cloud computing}\label{artcom}

Given the speed of development in the ICT area, it seems feasible to realize something like a ``Living Earth Simulator'' to simulate the entire globe, including all the diverse interactions of social systems and of the economy with our environment.
However, this requires to solve a number of problems, including the following ones:

\begin{enumerate}
\item to provide suitable hardware systems,
\item to transform data into knowledge, 
\item to create suitable software solutions, 
\item to build up a larger social supercomputing community, and
\item to improve the scientific methodology and validation of social simulation.
\end{enumerate}

\subsection{Europe has to catch up in the area of supercomputing and data centers}

Today, challenges such as global climate simulation require the use of the fastest supercomputers of the world. Simulating life on Earth and everything it relates to will require even bigger computer power. Such planetary-scale simulations will be unprecedented in their scale and level of detail. The situation is similar for  the collection and processing of the relevant data (see Ref. \cite{VisioneerCrisis}). The potential social and economic benefits of social super-computing would be enormous, from the prevention or mitigation of future crises up to the creation of new business opportunities (see Ref. \cite{VisioneerInnovationAccelerator}). 
\par
An analysis of the list of top 100 supercomputers in the world, however, shows that Europe is lacking behind in terms of computer power (see Appendix \ref{top100}). While it appears appropriate that 19 out of the 50 fastest supercomputers are located in Europe, 20 in the United States, and 9 in China or Japan, the top 20 list looks concerning: Only 5 of them are located in Europe, 4 in China or Japan, but 11 in the United States. The top 10 list is even alarming: Only 1 of the 10 fastest computers in the world is in Europe, 2 in China and Japan, while 7 of them are in the USA. The PRACE project \cite{PRACE} is trying to address Europe's computational gap on the long term by supplying exa-scale computing power by 2019.

A similar picture is found for data centers, the creation and maintenance of which poses increasing costs and problems, considering in particular the rapidly growing data generation rates and the difficulties in integrating multiple concurrent data-streams and data from offline repositories. This will require entirely new approaches to data processing. Today, most data are collected and stored by large Internet providers, government agencies, and private companies, while public research institutions are far behind. This situation implies a number of problems, including the lack of oversight of what is actually being done in different parts of the system, leading to multiple investments, while remaining knowledge gaps are not efficiently filled  \cite{VisioneerInnovationAccelerator}. Evaluating the diverse digital traces that human activities leave behind, this deficiency could be overcome.

\subsection{Turning data into knowledge: Future challenges to be addressed}

Current computer-based approaches in the social sciences still have a number of serious limitations which must and can be overcome, as discussed in the following:
\begin{itemize}

\item The discussion around the simulation of Volcanic ashes shows that it is not sufficient to have computer simulations. One rather needs a concept integrating real-time measurement and monitoring, e.g. through suitable sensor networks or other data sources. The goal is to reach zero-delay data mining which allows one (i) to feed boundary conditions and interaction network data into large-scale computer simulations, (ii) to calibrate parameters and determine advance warning signs on-the-fly, (iii) to discover characteristic patterns (``stylized facts''). 

\item Today's simulations are still restricted in scale and level of detail, and they lack integration. When performing transport-related simulations, for example, the time scales of traffic control, road repair, transport planning and regional planning range from seconds to 30 years. Such multi-scale simulations are still not operational on the level of a single region, and even less on a global scale.

\item Different sectors of the techno-socio-economic-environmental system need to be computed in an integrated way, not in separation. Today, we are still lacking a sufficient fiancial market simulation as well as an integrated micro-macro-simulation of the economic system. While this goal may be reachable to a certain extend within a couple of years, an integration with models of the environment and social system appear necessary as well. It is clear that the spreading of diseases is influenced by human transport and impacts economics and social behavior. In a similar way environmental changes affect agriculture, forestry, and other economic sectors, which may cause migration, integration problems, and conflicts or even wars. Therefore, it is crucial to gain an integrated systemic view. 

Causality networks \cite{Helbing_Ammoser} and systems dynamics approaches \cite{sterman00a} can certainly be useful to explore possible cascading and feedback effects. However, they are quite limited in scope, as they do not sufficiently consider the variability in space and of other factors, and oversimplify the interactions in the system. This could be overcome when computer models are combined with massive data mining.

\item The behavioral response to information about the system needs better modeling. This has, for example, become obvious for  predictions of epidemic models, where disease spreading rates change through behavioral adaptations. We also need to learn more about collective behavior, coalition formation, political decision making, and lobbying. All this requires the integration of behavioral models and, in particular of psychological models, into simulations of socio-economic systems. In other words, the approach must go beyond rational agent approaches and reflect emotions, perception biases, and possibly other factors as well. Again, a data driven model promises progress, but needs to be combined with empirical and experimental approaches (in the lab and Web). 

Furthermore, we are currently lacking something like a ``social information theory'' which would allow one to assess the relevance and impact of certain kinds of information on socio-economic systems, or emergent effects that may result when two or more pieces of information are perceived in correlation. 
\end{itemize}

\subsection{Crafting new software tools}\label{newSW}

The analysis of massive amounts of data is already a great computational
challenge. The even bigger goal, namely to simulate society and the global economy as a whole,
must be approached in a structured way. Herein,  the modeling of human activities
will play an eminent role. Today, there are already a number of multi-agent simulation platforms for specific domains like traffic (e.g. Transims \cite{Transims}, Matsims \cite{Matsims}), economics (e.g. Eurace \cite{EURACE}) and finance (e.g. Santa-Fe Market Simulator \cite{ASM}, U-mart \cite{U-Mart}). However, they are still lacking crucial functionalities, including the integration of real-time data streams. Therefore, a wider approach to simulation is needed, which overcomes the boundaries between simulation and reality. For example, the increasing popularity
of communal on-line games could be used as an opportunity for policy and systems design (see Sec. \ref{SysDes}). 
\par
Considering the potentials and limitations of current state-of-the-art large-scale simulators, a global-scale super-simulator (``Living Earth Simulator'') would require to have features such as the following:
\begin{itemize}
\item massive computational resources (in order to reach a sufficient degree of realism),
\item high flexibility, robustness, and evolutionary, self-organizing (self-*) properties,
\item interconnectedness with a variety of other simulation tools for the sake of multi-scale micro-macro modeling,
\item interoperability with large-scale data sets (both regarding the use of data during simulations and their deposition 
in massive repositories),
\item support of the simultaneous integration of multiple data-sources, recombining real-time streams of information with historical datasets, 
\item measurement of boundary conditions and the properties of the relevant interaction networks ``on the fly'', 
\item fusion of data-driven and model-based components in real-time,
\item self-calibration during the simulation, as more and more data become available over time 
(e.g. determination of weather-dependent road capacities or of spreading rates of diseases), 
\item establishment of coherent, flexible and practical standards for interoperability and systems integration,
\item verification and validation through lab and Web experiments, and/or serious multi-player games,
\item equipment with powerful visual data exploration tools (see Sec. \ref{dataviz}).
\end{itemize}
Each of the above points involves fundamental challenges requiring to significantly advance today's ICT systems. In the following, we can just highlight a few points. For example, for the purpose of modeling complex systems (such as technical, social, economic and environmental systems and their interactions), it would be favorable to develop a software system which allows one to (semi-)automatically perform the following tasks:
\begin{itemize}
\item identify well measurable, interpretable and relevant variables in large data sets,
\item detect significant patterns in high dimensional complex systems and visualize them, 
\item extract a variety of mathematical laws consistent with available data sets and identify their crucial assumptions, approximations, and simplifications, 
\item assess the suitability to interpret the alternative models, to perform analytical studies, and to calibrate them,
\item evaluate their validity, 
\item support the process of interpreting the models,
\item help to distinguish descriptive from explanatory models and to find models of the latter type (providing a better level of understanding of causal relationships and systemic interdependencies),
\item perform sensitivity analyses with respect to parameter variations and model assumptions (e.g. structural variations),
\item perform analyses of the network robustness (e.g. for critical infrastructures such as transportation systems, power grids, the financial system, communication systems, the Internet etc.),
\item perform analyses regarding spatial dependencies,
\item identify early warning signs of upcoming critical states and systemic shifts through scenario analyses,
\item generalize detected patterns and simplify models for them to determine ``stylized facts'' of the studied system,
\item customize the degree of detail to the modeling purpose,
\item determine implications which distinguish alternative models, and derive procedures to test the model variants by complementary data analyses or experiments, 
\item identify, run, and visualize relevant simulation scenarios, and support their interpretation, 
\item determine the range of validity of the alternative models and the levels of uncertainty of the simulation scenarios studied. 
\end{itemize}
Such a new software platform would play a fundamental role in integrating
\begin{enumerate}
\item data generated from various sources, as described in Ref. \cite{VisioneerCrisis},
\item models from economics, the social sciences and other disciplines, 
\item extending these models into new directions, for example as discussed in Ref. \cite{VisioneerGrandChallenges}, taking into account the massive data provided by new powerful ICT systems.
\end{enumerate}
Besides, it appears desirable 
\begin{itemize}
\item to create distributed, adaptive processing and storage systems based on principles of collective intelligence,
\item to employ emergent computing and artificial immune systems,
\item to make swarm architectures programmable like single systems,
\item to create ICT platforms that can reconfigure themselves as well as adaptive algorithms that re-program themselves, allowing for changing interaction rules,
\item to develop a standardized framework for modelling agent interactions,
\item to consider cognitive and emotional factors supporting swarm intelligence, and
\item to enable ``cultural evolution'' in computer networks.
\end{itemize} 

\subsection{Building up a larger social supercomputing community} 

It is quite obvious that additional research, development, consultancy and education capacities must be built up in the areas of social simulation and integrative systems design in order to address the challenges of humanity more successfully in the future. Experiences from the establishment of the bioinformatics community or the climate and earth science communities may be helpful to consider, here.
\par
One way of extending these capacities is to connect the social
simulation community (e.g. the European Social Simulation
Association, ESSA \cite{ESSA}) with natural and computer scientists, engineers, and ICT researchers. This would pull in people and knowledge from a variety of relevant areas and make sure that the wisdom gained in the socio-economic sciences will be considered. It is clear that this requires to foster a multi-disciplinary interaction, which could be done by joint think tanks, summer schools, workshops, symposia, conferences, etc., joint research projects and publications, and the creation of a joint study direction, which may be called ``Integrative Systems Design'' or just ``Systems Science'' (see Sec. \ref{SysDes}). Moreover, the field and its progress could be largely stimulated by ``Hilbert Workshops'' (identifying grand scientific challenges), suitable incentive schemes, and tools that make communication and collaboration easier and more efficient (see Ref. \cite{VisioneerInnovationAccelerator}). 

\subsubsection{User-friendly tools for non-experts}

An additional route to take is the provision of better tools to social scientists and economists for the purpose of computer simulations. In the past, programming computer code has traditionally not been a focus in these areas, in contrast to the natural and engineering sciences. As a consequence, there is a huge gap between what is done in the academic world and in the governance and business world. This situation must be overcome by creating user-oriented ICT systems. For this, the following seems to be necessary:
\begin{enumerate}
\item to create ``non-expert systems'', which make the knowledge of humanity, simulation tools etc. accessible for everyone,
\item to design models for adaptive autonomous systems that interact with humans and social structures: ICT systems that think like humans, understand what they want, and adapt to them, 
\item to develop ICT systems that are intuitive, easy to use, customizable, reliable (self-repairing), flexible (self-adaptive), trustable (in particular self-protecting), sustainable, scalable, and interoperable (like plug-ins or apps or widgets, see section \ref{WoM}),
\item to facilitate an interface-free interaction with users (along the line of systems that can be instructed by voice or latest-generation computer games, which ``understand'' body language and mimics). 
\end{enumerate}

\subsubsection{Building up the ``World of Modeling'' (WoM), a toolbox of interoperable apps}\label{WoM}

User-friendly simulation tools would make it easier to perform research that is driven
by the problems that need to be solved rather than by existing theories. Today, researchers still have to spend too much time on programming their models, which implies that these tend to be oversimplified. It seems already beyond the scope of a PhD thesis or the capability of a single research team to create a simulator that would allow to study the implications and differences of existing decision and behavioral models---not to mention more sophisticated models that may be developed in the future. For the way in which certain behavioral strategies (or opinions) spread, for example, numerous plausible, but mutually inconsistent approaches have been proposed, including the following ones \cite{FutureofSocialExperimenting}:

\begin{itemize}
\item strategic optimization considering likely future decisions of all involved parties, assuming that everybody else would behave so as well,
\item ``satisficing'', i.e. the application of aspiration-dependent rules (trying to reach a certain minimum performance), 
\item reinforcement learning, 
\item unconditional imitation of the best performing neighbor or interaction partner,
\item best response in the next interaction, given the current situation, 
\item multi-stage strategies such as tit for tat, 
\item the application of stimulus-response rules such as win-stay-lose-shift, 
\item the use of probabilistic rules such as the proportional imitation rule, the Fermi rule, or the unconditional imitation rule with a superimposed randomness (``noise''),
\item rules based on social influence (such as majority rules), giving rise to herding effects.
\end{itemize}

Despite all of these rules (and sometimes several variants of them) have been formalized and have testable implications, their range of validity has still not been experimentally determined. Most likely, different individuals apply different rules, and the application of the respective behavioral choice rule may even be situation-dependent (``framing effect'') \cite{Tversky/Kahneman:1981:Framing}. 

Complementary, one would also have to test the impact of different interaction networks and different initial conditions or even ``history dependencies'', to get a sufficiently differentiated picture of socio-economic reality (or possible realities) through computer simulations. Note that, at least to a certain extent, such a simulation approach would allow one to identify sensitivities to parameter changes, and even structural instabilities, which may cause unexpected systemic shifts (``phase transitions''). 
\par
To assess the sensitivity or robustness of socio-economic simulation results with regard to the underlying assumptions, it will be necessary to simulate a large variety of interaction rules and settings, and to compare their implications. This should be doable by simply ``plugging in'' model components when setting up the model underlying a computer simulation. Going one step further, the ICT system could also create a number of scenarios in a (semi-)automated way. This multiple-world-view approach would be compatible with the ``possibilitistic'' and ``pluralistic modeling'' strategies, which seem to be most appropriate for the analysis of socio-economic systems \cite{PluralisticModeling}.
\par
To build up the simulation capacities we are calling for, the scientific community would have to create standardized app-like model components, which are compatible and interoperable. Successful examples of such an approach are Web browsers add-ons (e.g. see \cite{Firefox_addons}), modules of content management systems (CMS) (e.g. see \cite{Drupal_Modules}) or the famous apps for mobile phone operating systems (see e.g. \cite{Apple_apps} or \cite{Google_apps}). The way in which open communities such as Linux or Wikipedia are organized can give a good orientation of how to do this (see also Ref. \cite{VisioneerInnovationAccelerator}). The result of such community-wide activities creating model components for the simulation of socio-economic systems would be the creation of a big library (``toolbox'') for agent-based and other simulation approaches, which could be called the ``World of Modeling'' (WoM). The approach would overcome the current situation, in which we have a pletora of limited and mutually incompatible simulation tools such as Swarm \cite{SWARM} or Repast \cite{Repast}, which provide low-level libraries, and graphical modelling environments such as Netlogo \cite{Netlogo} or Sesam \cite{Sesam}, which are more suited for beginners (for a comprehensive comparison table of the features of available packages for agent-based simulations see Ref. \cite{Wikipedia_Comparison_ABM}). Even powerful simulation packages like MatLab \cite{Matlab} are not suited for massive agent-based simulations or anything that would come close to the envisaged ``Living Earth Simulator''. Moreover, there is still a lack of analytical tools capable of exploring ABM results from a scientific point of view.

\section{Proposals to establish centers for integrative systems
  design}\label{SysDes}

Besides analyzing, modeling and simulating our world, additional efforts will also be required in the area of managing complexity and integrative systems design in order to be able to prevent, mitigate, or cope with crises more efficiently.

\subsection{Predictability in socio-economic systems: Possibilities and limitations}

Many people believe that, in order to create a crisis-relief system, one needs to be able to predict the future of the system, but doubt that such prediction is possible for socio-economic systems. However, both is not exactly true: 
\begin{enumerate}
\item There are many other purposes of modeling and simulation besides prediction \cite{epstein2008}. 
\item ``Model predictions'' should be better understood as ``model implications'' rather than ``model forecasts'', e.g. a statement that says what systemic outcome is expected to occur for certain (regulatory) boundary conditions, e.g. cooperation, free-riding, or conflict \cite{HelbingJohanssonNorms,10.1371/journal.pcbi.1000758}.  
\item It is important to point out that forecasts are possible \textit{sometimes}. A famous example is Moore's law regarding the performance of computer chips.  Moreover, while the detailed ups and downs of stock markets are  hard to predict, the manipulation of interest rates by central banks leads, to a certain extent, to foreseeable consequences. Also the sequence in which the real-estate market in the US affected the banking system, the US economy, and the world economy was quite logical. That is, even though the exact timing can often not be predicted, causality networks allow one to determine likely courses of events, which in principle enables us to take counter-actions in order to avoid or mitigate the further spreading of the crisis \cite{Helbing_Ammoser}.
\item As weather forecasts show, even unreliable short-term forecasts can be useful and of great economic value (e.g. for agriculture). Another example illustrating this is a novel self-control principle for urban traffic flows, which was recently invented \cite{Helbing_Laemmer_selfcontrol,Helbing_Laemmer_Patent}. 
Although its anticipation of arriving platoons is of very short-term nature, it manages to reduce travel times, fuel consumption, and vehicle emissions.
 
\end{enumerate}
In conclusion, prediction is limited in socio-economic systems, but more powerful than is usually believed. Moreover, in certain contexts, it is not necessary to forecast the course of events. For example, in order to reduce problems resulting from bubbles in financial markets, it is not necessary to predict the time of their bursting or even to know about their existence. What is required is a mechanism that destroys any forming bubbles, as long as they are small. Such mechanisms are possible. For example, they could be implemented by introducing additional sources of randomness into the system \cite{HelbingSystemicRisks_SE,NewspaperArticleWithMarkusChristen2010}.

\subsection{Paradigm shift from controlling systems to managing complexity}\label{managing_complexity}

The idea to \textit{control} socio-economic systems is not only inadequate---it is also not working well. Socio-economic systems are complex systems. Therefore, cause and effect are usually not proportional to each other. In many cases, complex systems tend to resist manipulation attempts, while close to so-called ``tipping points'' (or ``critical points''), unexpected ``regime shifts'' (``phase transitions'', ``catastrophes'') may happen. Consequently, complex systems cannot be controlled like a technical system (such as a car) \cite{citeulike:3975610}.

Note that the above property of systemic resistance is actually a result of the fact that complex systems often self-organize, and that their behavior is robust to not-too-large perturbations. While forcing complex systems tends to be expensive (in case systemic resistance is strong) or dangerous (in case an unexpected systemic shift is caused), the alternative to support the self-organization of the system makes much more sense. Such an approach ``goes with the flow'' (using the natural tendencies in the system) and is resource-efficient. Therefore, a reasonable way to manage complexity is to guide self-organization or facilitating coordination \cite{citeulike:3975610,citeulike:6622211}.

In a certain sense, this self-organization or self-control approach moves away from classical regulation to mechanism design. Regulation often corresponds to changing the boundary conditions, while mechanism design changes the interactions in the system in a way that reduces instabilities (e.g. due to delays) and avoids that the system is trapped in local optima (and continues its evolution to a system-optimal state). For example, slightly modifying the interaction of cars by special driver assistant systems can stabilize traffic flows and avoid bottleneck effects to a certain degree \cite{KestingTreiberHelbingTRC}. 

\subsection{Self-defeating prophecies and proper design of information systems}

It is often pointed out that socio-economic systems would not be predictable, because the reaction of people to information about the system would destroy the validity of the forecast. If done in the wrong way, this is actually true. Let us illustrate this by an example: Assume that all drivers are given the same information about existing traffic jams. Then, drivers may most likely over-react, i.e. more drivers may use an alternative road than required to reach a system-optimal distribution of traffic flows \cite{HelbingInSchreckenbergSelten}. 
\par
However, as has been shown in laboratory route choice experiments, an almost optimal route choice behavior may be reached by an information system that gives user-specific recommendations. In other words, some drivers would be asked to stay on the road, and others to leave it. By introducing a form of compensation for the fact that not everyone follows the recommendations, one can avoid over- or under-reactions of drivers in congestion scenarios \cite{HelbingInSchreckenbergSelten}. 
\par
One crucial issue of such individualized recommender systems, however, is their reliability. An unreliable system or one that is systematically biased, will be only poorly followed, and people quickly compensate for biases. Therefore, it is also essential to design the information system in a way that is fair to everyone. That is, nobody should have a \textit{systematic} advantage. Nevertheless, the system should be flexible enough to allow a trading of temporary (dis)advantages. For example, somebody who was asked to take the slower road on a given day (and who would have a right to use the faster road on another day), may still use the faster road. However, he or she would have to pay a fee for this, which would be earned by somebody else who would exchange his or her ``ticket'' for the faster road for a ``ticket'' for the slower road (such that the system optimum state would still be maintained)  \cite{HelbingInSchreckenbergSelten}. 
\par
In summary, the above described information system would not cheat anyone, and it would be flexible and fair. Only people who use the faster road more often than average would have to pay a road usage fee on average. A normal driver would either pay nothing (when following the user-specific recommendation) or pay on some days (when being under a pressure of time, while the recommendation asks to take the slower road). However, the same amount of money could be earned on other days by taking the slower road. In other words, fair usage would be free of charge on average, and drivers would still have a freedom of route choice. The primary goal of the system would not be to suppress traffic flows through road pricing, but the pricing scheme would serve to reach a system optimal traffic state.

\subsection{New approaches and designs to manage complexity}

There is no single scheme, which allows to manage all complex systems optimally, independently of their nature. The success of a management concept very much depends on the characteristics of the system, e.g. its degree of predictability. The systems design must account for this. 
\par
In case of long-term predictability, there must obviously be an almost deterministic relationship between input- and output-variables, which allow one to change the temporal development and final outcome of the system. If long-term predictability is not given, management attempts must be oriented at a sufficiently frequent re-adjustment, which requires a suitable monitoring of the system. 
\par
As the example of weather forecasts shows, even unreliable short-term predictability can be very useful and economically relevant (e.g. for agriculture).\footnote{Another example is the ``self-control'' of urban traffic flows, which is based on a special, traffic-reponsive kind of decentralized traffic light control \cite{Helbing_Laemmer_selfcontrol}, see Sec. 
\ref{managing_complexity}}  The success principle in case of short-term forecasts is the flexible adjustment to the local environment, while well predictable systems often perform well with a fixed (rather than variable) organization principle. 
\par
Even in case of a system that cannot be forecast over time at all, it is possible to modify the system behavior. Taking the time-dependent course of the stock markets as an example, introducing a Tobin tax would reduce excessive levels of speculation. Moreover, introducing ``noise'' (further sources of unpredictability), would destroy undesirable correlations and impede insider trading \cite{NewspaperArticleWithMarkusChristen2010}.
\par
These are just a few examples illustrating that there actually {\it are} possibilities to influence systems involving human behavior in a favorable way. A more detailed discussion of the issue of managing complexity is given in Refs \cite{citeulike:3975610,citeulike:608190,Fabien_2008,citeulike:5492267,Fradkov99nonlinearand}.

Generally, there are two ways of influencing the dynamics and outcome of a system by changing the ``rules of the game''.
If the interaction in the system are weak, the system dynamics can be well influenced by modifying the boundary conditions of the system (i.e. by regulatory measures). However, if the interactions are strong, as in many social and economic processes, the self-organization of the system dominates the external influence. In this case, a modification of interactions in the system (mechanism design) seems to be more promising (as in the traffic assistance system that reduces the likelihood of congestion by special driver assistance systems \cite{KestingTreiberHelbingTRC}). Of course, both measures may also be combined with each other.
\par
While mechanism design is relatively common in computer science and some other areas (e.g. in evolutionary game theory, mathematics, and partly in physics), it seems that these methods have not been extensively applied in the socio-economic sciences so far. For example, there are many different mechanisms to match supply and demand, and it would be interesting to know what systemic properties they imply (such as the level of stability, the efficiency of markets, the resulting wealth distribution, the creation of investment opportunities, etc.). There is also an on-going debate about the best way to involve the population in the political decision-making process. It is still not clear how to reach the best combination of top-down and bottom-up elements of interaction, and how to find the best balance between centralized and decentralized coordination approaches. All this poses interesting and practically relevant challenges that determine the prosperity and well-being of societies (see also Ref. \cite{VisioneerInnovationAccelerator}).
\par
Moreover, in the past, mechanism design has been applied basically to (sub)systems (i.e. parts) of the complex overall system we are living in. However, due to the interconnection of all sorts of (sub-)systems (e.g. of the traffic, supply, industrial, and environmental systems), measures in one (sub)system may have undesired side effects on other (sub)systems. In fact, for fundamental reasons, it is quite frequent that taking the best action in one (sub)system will {\it usually} negatively affect another (sub)system, such that an optimal state of society is not reached (setting aside for the moment the practical problem of what the optimal state of society would be). The poor performance of traffic light controls that optimize locally (without a coordination with neighboring intersections) is a good example for this \cite{KestingTreiberHelbingTRC}. In networked systems, undesirable feedback effects (such as spill-over effects in urban traffic flows) are quite common.

\subsection{Need of multi-disciplinary collaboration}

As a consequence of the complexity of socio-economic systems, one requires an integrative approach that comprises technical, social, economic {\it and} environmental systems and their interactions. A good concept to manage complexity must be integrative and holistic in nature. Such an approach may be called ``integrative systems design''. Extending the idea of a ``policy wind tunnel'' (a computer-based, experimental device to support decision-making) to large, global-scale systems, this will certainly require massive computer power and huge amounts of data, as envisaged by the concept of the Living Earth Simulator (see Sec.\ref{newSW}). Moreover, in order to get beyond descriptive (fit) models in favor of explanatory models, it would call for the integration of a large variety of competencies, including

\begin{itemize}
\item economics,
\item social sciences (also considering psychology, anthropology, history, \ldots),
\item statistical physics,
\item the theory of complex adaptive systems (including self-organization, emergence),
\item network theory,
\item cybernetics, etc. 
\end{itemize}

This implies the necessity of a multi-disciplinary approach. Due to the difficulties related to the fact that multi-disciplinary research, education and funding does not fit well into existing institutional settings, it would be advised to build up a new, multi-disciplinary field and study direction. This would involve the above subjects, and also computer science (including programming, visualization, AI, robotics, etc.), mathematics, ecology, and biology (organization of social species, the immune and neural systems, etc.). It is quite obvious that there is a lack of experts who would oversee all these areas and would be capable of an integrative systems design approach. Such experts overlooking a broad range of problems and methods would be very much needed in engineering, economics, and areas involving decision-making about complex issues (such as politics). 

\subsection{New ICT tools for integrative systems design}\label{shift1}

In the past, it was impossible to experiment with our future. This made social sciences different from the natural and engineering sciences, in which different options can be tried out before choosing one. In the future, we will also be able to make experiments with different socio-economic designs (e.g. various market mechanisms). Recently, besides the classical method of role plays, lab experiments are establishing as a new scientific method in the socio-economic sciences \cite{citeulike:191352,ExpEco_Rethinking,EcoLabIntesiveCourse,friedman1994emp,RePEc:cup:cbooks:9780521618618}. These approaches are now complemented by Web experiments \cite{citeulike:520717} and studies of multi-player on line games (such as ``Second Life'') \cite{FutureofSocialExperimenting,szell:dynamics,citeulike:1507451,johnson09}.

In other words, so-called serious games, which represent features of society and economics in a quite realistic way, may be used to gain insights into human behavior and effects of social or economic interactions on the systemic level. Such virtual realities allow one, in principle, to explore a number of different scenarios and worlds \cite{citeulike:520717}. For example, a variety of designs of a new shopping mall, airport, railway station, or city center, could be populated in virtual reality and tested by their future users. In this way, one can discover weaknesses of designs in advance and identify the most popular architecture or city design. 
\par
It is obvious that similar approaches can be applied to a variety of different institutional design problems. Using serious multi-player games complementary to lab, Web and real-life experiments (as far as these is ethically opportune) can create significant new insights and reveal weaknesses of computer models, such as over-simplified model assumptions or overlooked influence factors. Therefore, they are a necessary instrument in the ICT tool set needed to overcome our knowledge gaps regarding complex socio-economic systems.
\par
As a result of these new ICT-based opportunities, there will be a paradigm shift in socio-economic decision-making.
In the past, it was common to implement suitably appearing new measures based on experience and intuition. However,  many decisions today have to target problems for which no previous experience is available, and the complexity is so high that feedback and side effects of new measures cannot be anticipated well enough by human intuition anymore.

\subsection{Paradigm shift in decision-making}\label{shift2}

Therefore, it seems likely that the future way of implementing new measures will look as follows: Before the implementation, large-scale computer simulations, fed by huge datasets, will explore different possible scenarios and their side effects. Complementary, one will perform experiments in the lab, Web, and virtual realities. All these experimental techniques (including the computer experiments) will serve to gain in advance the best possible knowledge of 
the decision options and their likely implications for the concerned individuals and the socio-economic system. For example, one could study stress test scenarios for financial systems under as-realistic-as-possible conditions, involving interactions between all stakeholders (banks, governments, customers, tax payers, ...), while today's stress test still largely neglect interaction effects. Based on the so collected experience, the best option would be chosen and implemented by the decision-makers, considering priorities, normative, and ethical issues. 

Furthermore, it would make sense to implement the best measures (institutional designs) in limited parts of the system first and evaluate their performance. After this practical test (``pilot studies''), the best solution would be transferred to the whole system, but the success of the measure would be continuously monitored. In some sense, this approach to implementing innovations is more along the lines how nature seems to work. In fact, the described approach basically follows the principles of evolution, with the main difference that some of the testing of new solutions happens in the virtual rather than the real world, and only the best variants are deployed in reality. It is obvious that, in the creation and selection of new solutions, evolutionary algorithms \cite{fogel2000introduction,fogel2006evolutionary} will play a major role. Besides, one can probably learn from complex adaptive systems such as neuronal networks \cite{haykin1994neural,hagan1996neural}, immune systems \cite{janeway2001immunobiology}, or the organization of ecological systems \cite{citeulike:3115948,MayAND:2007}.  

\subsection{The importance of visualization and decision arenas}\label{darenas}

Another crucial element of this integrative systems design concept is visualization. It matters not only for virtual reality applications. Also large-scale data mining and massive computer simulation imply a need for new visualization techniques and approaches. This subject is addressed in detail in Appendix \ref{dataviz}. Besides new visualization approaches, one will also require new visualization centers for sophisticated three-dimensional animations and the demonstration to a larger number of decision-makers, from expert panels, over managers and politicians to the interested public.
\par
Combining suitable data, models and visualization tools, the final goal would be to develop a \textit{virtual observatory} of social, economic and financial systems with the possibility to explore new scenarios and system designs through what it sometimes called a ``policy wind tunnel''. The corresponding ICT systems should be suited to craft policy options, contingency plans, and intelligent responses to emerging challenges and risks. They will allow one to develop concepts and institutional designs for a flexible, adaptive, and resource-efficient real-time management of complex socio-economic systems and organizations. Decision arenas visualizing, animating, and illustrating the acquired scientific insights will provide valuable decision-support to decision-makers, as they will make counter-intuitive feedback, cascading and side effects understandable. 

Decision Arenas will also be useful to craft new designs for smarter cities, transport, traffic and logistics, and for intelligent energy production and consumption, as well as new institutional frameworks and settings to support social and economic participation.

\subsection{Ethical Issues}\label{Ethical}

It should finally be pointed out that manipulating social or economic systems potentially raises ethical issues. People working in this area should, therefore, follow an ethical codex which could be imagined similar to Hippocratic Oath of doctors or other vows which as required from certain public officers.

The following goals seem to be widely acceptable:
\begin{itemize}
\item promote human well-being,
\item increase the self-awareness of society,
\item reduce vulnerability and risk, 
\item increase resilience (the ability to absorb societal, economic, or environmental shocks), 
\item avoid loss of control (sudden, large and unexpected systemic shifts),
\item develop contingency plans, 
\item explore options for future challenges and opportunities,
\item increase sustainability,
\item facilitate flexible adaptation,
\item promote fairness, 
\item increase social capital and the happiness of people,
\item support social, economic and political inclusion and participation,
\item balance between central and decentral (global and local) control,
\item protect privacy and other human rights, pluralism and socio-bio-diversity,
\item support collaborative forms of competition and vice versa (``coopetition'').
\end{itemize}
Unethical use of powerful ICT tools should be forbidden. From people working with them, one should require a minimum level of qualification and training, including ethical issues (``licence''). The activities of these ICT systems may need to be recorded and monitored in a way that is sufficiently transparent. The public should have a right to be informed about related activities (be they at academic or government institutions, or in private companies). Security measures will depend on the size of the simulations and application areas. The software should be designed in a way that it can be remotely deactivated by a number of specially authorized, independent people, if they agree on the necessity of such a step (see Ref. \cite{VisioneerCrisis} for details of such a mechanism). An ethical committee should accompany and supervise related activities.

\section{Summary}

Today, statistical and national planning offices are using a large variety of
models for different sectors of the economy and society, but these are still largely
disconnected from each other and hard to integrate, so that systemic interaction effects
are hard to assess. Considering financial and economic instabilities, conflict, large-scale environmental
changes, threats through diseases, cyber-crime, or problems resulting from demographic change (such as the instability of
social benefit systems), etc., our society needs to undertake any possible effort to get a better understanding
of these problems and possibilities to mitigate them. In this
context, the role of data is crucial to provide decision makers all relevant information 
needed to act in a timely manner and find long-term sustainable solutions. 
\par
Hence, there is a strong need for integrated tools that are able to close the gap between mainstream 
linear fit models and the required dynamic and holistic thinking that is appropriate to address the global
challenges of the twenty-first century. Systemic and dynamic approaches have
the advantage to consider feedback loops and to be able to reveal
previously hidden consequences, while also supporting the assessment
of the effectiveness of policy measures. Massive agent-based simulations seem to be particularly
promising in this context, as they allow one to handle many different kinds of agents (individuals, groups, companies, etc.), the variability of their individual properties, and interactions among the agents. Ideally, in the future, the implementation of such agents should reflect cognitive features and emotions, behavioral adaptation, the response to information, and the tendency to show collective behaviors (e.g. to form alliances or groups).
\par
The final goal of such super-computer-based activities 
is to monitor and improve the performance of the whole Living Earth system (including its 
socio-economic life), to assure systemic resilience and avoid 
to run into optimization traps.\footnote{In fact, in certain systems (such as various transport, logistic, or production
systems), optimization tends to drive the system towards instability,
since the point of maximum efficiency is often in the neighborhood or
even identical with the point of breakdown of performance. Such
breakdowns in capacity or performance can result from inefficiencies
due to dynamic interaction effects. For example, when traffic flow
reaches its maximum capacity, sooner or later it breaks down. As a
consequence, the road capacity tends to drop during the time period
where it is most urgently needed, namely during the rush hour \cite{Helbing:1117341,Helbing:480450}.} Note that na\"{\i}ve optimization tends to eliminate heterogeneity \cite{EconomistFailed} and redundancies
in the system and may increase the vulnerability to perturbations, i.e. decrease robustness and resilience. 
Moreover, there exists special ``evolutionary dead ends'', in which
gradual optimization simply does not work. In such cases, the best
solution may be the combination of two bad solutions. Therefore, it may
be easily overlooked. These dangers of na\"{\i}ve optimization suggest to pursue a different approach oriented at the principles of evolution, i.e. to explore new solutions locally and test them, and if they perform well, then disseminate and apply them to larger areas (see Sec. \ref{shift2}). 
\par
For all these reasons, there is a strong need to set-up a network of Centers for integrative systems design, capable of gaining an ``ecological perspective'' of all relevant interactions in socio-economic-techno-environmental systems. Within this holistic approach, these centers shall be able to run all potentially relevant scenarios, identify causality chains, explore feedback and cascading effects for a number of model variants, and determine the reliability of their implications (given the validity of the underlying models). These Centers will compute, analyze and use independent information to address the impact of model assumptions and reveal hidden constraints. They will be able to detect possible negative side effect of policy decisions, before they occur. The Centers belonging to this network of Integrative Systems Design Centers would be focused on a particular field, but they would be part of an attempt to eventually cover all relevant areas of society and economy and integrate them within a ``Living Earth Simulator''. 
\par
The results of all research activities of such Centers would be turned into informative input for 
political \textit{Decision Arenas}. For example, Crisis Observatories \cite{VisioneerCrisis} (for financial instabilities, shortages of resources, environmental change, conflict, spreading of diseases, etc.) would be connected with such Decision Arenas for the purpose of visualization, in order to make complex interdependencies understandable to scientists, decision-makers, and the general public. By informing all the stakeholders in a better way, it will be easier to assess different options and their often counter-intuitive feedback, cascading, and side effects. In this way, it will also become more efficient to develop contingency plans and intelligent responses to emerging risks or opportunities, having the best of all knowledge at hand. The tight link between Integrative System Design Centers and Decision Arenas will also foster the development of concepts and institutional designs for a flexible, adaptive, and resource-efficient real-time management of complex socio-economic systems and organizations. One of the strategies to explore new system designs will follow an evolutionary approach based on techniques such as metaheuristic optimization (genetic algorithms, swarm optimization, etc.). It will also profit from evolutionary computing methods such as differential evolution, learning classifiers, artificial immune systems, etc. In such a way, new designs will be determined for smarter cities, transport, traffic and logistics, for intelligent energy
production and consumption, or for a better social and economic participation.

\section*{Acknowledgements}

The authors are grateful for financial support by the Future and Emerging Technologies programme FP7-COSI-ICT of the European Commission through the project Visioneer (grant no.: 248438).
They would also like to thank for stimulating discussions, feedback on the
manuscript and contributions to the Visioneer wiki: Ravi Bhavnani, Anna Carbone, Virginia Dignum, Andreas Flache, Mike Holcombe, Janos Kertesz, Imre Kondor, Rosario Nunzio Mantegna, Tamara Mihaljev, Dan Miodownik, Stefan Reimann, Xavier Rubio, Christian Schneider, David Sumpter, Piotr Swistak, Gabriele Tedeschi, Pietro Terna, and Arne Traulsen.

\appendix

\section{Visual problem solving}\label{dataviz}

The problem of how to find valuable information from (relatively)
large data sources is probably one of the most pressing in the history of
science. If human intuition alone may have sufficed to cope with the contents
of several books or with analysing two-dimensional and
time-series plots, today the exponential growth of data generation
rates imposes the use of more advanced visualizations techniques to
grasp the breadth and depth of the interrelations among
petabytes of data. 

Statistical and structural analyses are useful tools, but they come into play
later, when the first hypotheses have been made. At the
early stages of scientific analyses, when little is known about the
data, but also in its interpretation and maturation process, the intuition of
experts is very important and must enter the modeling process in a frictionless
and interactive way. It can largely be stimulated by the great processing capacity of
human visual system, which can easily notice correlations and
quickly recognize variations in colors, shapes,
movements, etc. Therefore, the visual exploration of data is crucial,
particularly for the hypothesis generation process
\cite{VisualDataMining}. Moreover, at more advanced stages of the work,
it can serve to quickly analyze the results of simulations and
experiments, e.g. regarding the correct fitting of parameters, and
it helps to check or reject hypotheses, at least on an intuitive
level.

Unfortunately, the development of novel techniques for the visualization
of scientific data did not catch up with the astounding improvements
in computational capacity over the last decades. This has created a
dangerous \textit{informational paradox}, where one may have to invest one
month to understand the results of a one-hour simulation job. As
Kageyama and Ohno have pointed out, computational science without
proper visualization tools is like ``a high-rise building
[not] supported by balanced
pillars''\cite{VirtualReality}. In fact, now more than ever, there is
an acute need of enhanced visualization tools. This urgency has not
been generated due to a surge in the size of datasets, but it was caused,
in particular, by the explosion of their dimensionality
aspect, i.e. the number of observations for each entity. In fact,
``what makes the data \textit{Big} is repeated observations over time
and/or space''\cite{PathologyBigData}. This high dimensionality has made
the standard visualization techniques inadequate, which were suitable for data
with low dimensionality, such as two- or three-D plots, OLAP (On Line
Analytical Processing), etc.

Herein, we will briefly review the state-of-the-art visualization techniques and describe how they should be integrated in a unique visualization tool that both scientists in integrative system design centers and policy-makers in decision arenas would enormously benefit from.

\subsection{Visual data exploration}

Fortunately, the new paradigm for future visualization tools
has been already suggested. It can be summarized by
three simple directives, constituting the \textit{Information Seeking
  Mantra}\cite{EyesHaveIt}. Namely: ``overview first, zoom and filter,
then details-on-demand''. In principle, a data scientist should be
able to easily execute the three above-mentioned steps, passing
seamlessly from one to another, without losing the overall
picture. It is important to note that the technology to implement
each step of the Information Seeking Mantra is already existing, but what
is still missing is their seamless integration. That requires
special attention for dynamic interaction and distortion
techniques. To this extent, interactive zooming should not display
only bigger objects, but also change the type of data representation
according to the scale, switching for example from colored pixels to
high-resolutions icons such as sticky figures \cite{IconsViz} or faces
\cite{FacesViz}, in order to exploit the great potential of the human
brain to deal with physiognomy. Moreover, filtering should be enabled
directly at the visualization level, by implementing mechanisms to
immediately and interactively process and display portions of the dataset in
different ways, and scan them by ``magnifying lenses'' \cite{MagicLenses}. One possible
transformation in presence of high-dimensionality is the
parallel coordinates technique \cite{ParallelCoordinates}, while
hierarchical structures can be rendered by rotating cone
trees\cite{ConeTrees}, although for particularly dense trees special
spatial navigation support is required. Alternatively, the use of
properties of non-Euclidean geometries can overcome this
problem \cite{HyperbolicTrees}. However, the most space-efficient
solutions are squarified and cushion treemaps
\cite{CushionTreeMaps,SquarifiedTreeMaps}, which are especially
effective to represent file system structures.

As just shown, many visualization technologies are already available,
but their implementation in a comprehensive tool capable to deal with
big amounts of data and available for the scientific community is still
missing. Probably the best results in this direction could come only from
a joint collaboration between visualization scientists and graphic
artists, designers, and communicators, as suggested also by a recent
Nature article \cite{DistillingMeaning}.

Finally, if the Information Seeking Mantra had been written more
recently, it would have probably entailed a fourth directive in order to
study data in a distributed way from different individual perspectives, namely
\textit{share}. The Web is in fact currently experiencing a democratic
data revolution where not only the datasets, but also the visualization tools
required to manipulate them are shared. Dynamic interaction with the data and Web 2.0 
social components are the staple features of such Web platforms (see Appendix of Visioneer White Paper ``From Social Data Mining to Forecasting Socio-Economic Crisis'' \cite{VisioneerCrisis}). Applications could be manifold. For example, people could
explore new architectural designs (such as airports, railway stations, pedestrian zones,
stadia or shopping malls) before they are built, as interactive virtual multi-player
environments would allow them to populate them in advance (see \cite{ScienceSim}). 

\subsection{Virtual Reality for Scientific Visualization}

Today, the frontier of human-computer interfaces is still called
Virtual Reality. Although it is not a new technology, it is still in
an maturating state. Nevertheless, nothing else at present supports
the understanding of complex high-dimensional data in
a similarly effective way.

In fact, virtual reality provides a rapid and intuitive way to
explore high data volumes with a rich set of spatial and depth
cues \cite{VirtualReality2}. There are many reasons, which make virtual
reality particularly appropriate for scientific visualization. First,
it can easily be oriented towards abstract and conceptual objects
which can easily go beyond traditional notions of data representation,
stimulating the scientists' intuition. Secondly, it allows a level of
control and direct interaction which can hardly be surpassed. Finally,
it has been proved it can be beneficial in debugging data acquisition
methods.

Notwithstanding its great potential, one of the reasons why it has
been only marginally adopted so far is the requirement of special, and often
costly, hardware and/or software (see \cite{VirtualRealityTrends}). Another
often-heard criticism is the lack of tools for specialized
visualization, or of an established framework to build such
applications on top of it \cite{VIRPI}.

Nonetheless, it is timely to focus research efforts in this direction,
because the potential benefits of breakthrough in this area would be
immense.

To this extent, it is worth mentioning the activities at the Earth
Simulation Center in Japan, where a team of scientists have been
working for years to develop a virtual reality solution for scientific
visualization called VFIVE \cite{VizEarthCenterAnnualRep}, the
source code of which is now publicly available \cite{VFive_site}.

\section{List of top 100 supercomputer centers}\label{top100}

Herein follows the list of the top hundred supercomputer sites in the
world. The list is updated to June 2010 and it is provided by Top500
\cite{Top500}, which uses the LINPACK benchmark\footnote{It measure
  how fast a computer solves a dense N by N system of linear equations
  $Ax = b$.} as an index to sort the sites by computing power.

\begin{center}

\begin{longtable}{ r p{12cm} l} \label{top500}
\\
\caption[World top 100 supercomputing centers ]{World top 100
  supercomputing centers \cite{Top500}.}
\\
\hline \multicolumn{1}{r}{\textbf{n.}} &
\multicolumn{1}{l}{\textbf{Supercomputing site}} &
\multicolumn{1}{l}{\textbf{Country}} \\ \hline
\endfirsthead

\multicolumn{3}{c}%
{{\bfseries \tablename\ \thetable{} -- continued from previous page}} \\
\hline  \multicolumn{1}{r}{\textbf{n.}} &
\multicolumn{1}{l}{\textbf{Supercomputing site}} &
\multicolumn{1}{l}{\textbf{Country}} \\ \hline
\endhead

\hline \multicolumn{3}{r}{{Continues on next page.}}
\endfoot

\hline
\endlastfoot

1 & Oak Ridge National Laboratory & United States \\ 
2 & National Supercomputing Centre in Shenzhen (NSCS) & China \\ 
3 & DOE/NNSA/LANL & United States \\ 
4 & National Institute for Computational Sciences/University of Tennessee & United States \\ 
5 & Forschungszentrum Juelich (FZJ) & Germany \\ 
6 & NASA/Ames Research Center/NAS & United States \\ 
7 & National SuperComputer Center in Tianjin/NUDT & China \\ 
8 & DOE/NNSA/LLNL & United States \\ 
9 & Argonne National Laboratory & United States \\ 
10 & Sandia National Laboratories / National Renewable Energy Laboratory & United States \\ 
11 & Texas Advanced Computing Center/Univ. of Texas & United States \\ 
12 & DOE/NNSA/LLNL & United States \\ 
13 & Moscow State University - Research Computing Center & Russia \\ 
14 & Forschungszentrum Juelich (FZJ) & Germany \\ 
15 & KISTI Supercomputing Center & Korea, South \\ 
16 & University of Edinburgh & United Kingdom \\ 
17 & NERSC/LBNL & United States \\ 
18 & Grand Equipement National de Calcul Intensif - Centre Informatique National de l'Enseignement SupÃ©rieur (GENCI-CINES) & France \\ 
19 & Institute of Process Engineering, Chinese Academy of Sciences & China \\ 
20 & Oak Ridge National Laboratory & United States \\ 
21 & Sandia National Laboratories & United States \\ 
22 & Japan Atomic Energy Agency (JAEA) & Japan \\ 
23 & King Abdullah University of Science and Technology & Saudi Arabia \\ 
24 & Shanghai Supercomputer Center & China \\ 
25 & Government & France \\ 
26 & University of Edinburgh & United Kingdom \\ 
27 & Swiss Scientific Computing Center (CSCS) & Switzerland \\ 
28 & SciNet/University of Toronto & Canada \\ 
29 & Government & United States \\ 
30 & ERDC DSRC & United States \\ 
31 & University of Colorado & United States \\ 
32 & New Mexico Computing Applications Center (NMCAC) & United States \\ 
33 & Computational Research Laboratories, TATA SONS & India \\ 
34 & Lawrence Livermore National Laboratory & United States \\ 
35 & DOE/NNSA/LANL & United States \\ 
36 & National Institute for Computational Sciences/University of Tennessee & United States \\ 
37 & Japan Agency for Marine -Earth Science and Technology & Japan \\ 
38 & IDRIS & France \\ 
39 & ECMWF & United Kingdom \\ 
40 & ECMWF & United Kingdom \\ 
41 & DKRZ - Deutsches Klimarechenzentrum & Germany \\ 
42 & JAXA & Japan \\ 
43 & US Army Research Laboratory (ARL) & United States \\ 
44 & Commissariat a l'Energie Atomique (CEA) & France \\ 
45 & Commissariat a l'Energie Atomique (CEA)/CCRT & France \\ 
46 & Joint Supercomputer Center & Russia \\ 
47 & HLRN at Universitaet Hannover / RRZN & Germany \\ 
48 & HLRN at ZIB/Konrad Zuse-Zentrum fuer Informationstechnik & Germany \\ 
49 & Total Exploration Production & France \\ 
50 & Government Agency & Sweden \\ 
51 & Computer Network Information Center, Chinese Academy of Science & China \\ 
52 & Lawrence Livermore National Laboratory & United States \\ 
53 & Information Technology Center, The University of Tokyo & Japan \\ 
54 & Kurchatov Institute Moscow & Russia \\ 
55 & Max-Planck-Gesellschaft MPI/IPP & Germany \\ 
56 & Institute of Physical and Chemical Res. (RIKEN) & Japan \\ 
57 & Pacific Northwest National Laboratory & United States \\ 
58 & EDF R\&D & France \\ 
59 & IT Service Provider & Germany \\ 
60 & IBM Thomas J. Watson Research Center & United States \\ 
61 & Naval Oceanographic Office - NAVO MSRC & United States \\ 
62 & Manufacturing Company & France \\ 
63 & Bull & France \\ 
64 & GSIC Center, Tokyo Institute of Technology & Japan \\ 
65 & University of Southern California & United States \\ 
66 & Lawrence Livermore National Laboratory & United States \\ 
67 & Stony Brook/BNL, New York Center for Computational Sciences & United States \\ 
68 & Maui High-Performance Computing Center (MHPCC) & United States \\ 
69 & Naval Oceanographic Office - NAVO MSRC & United States \\ 
70 & CINECA & Italy \\ 
71 & Center for Computational Sciences, University of Tsukuba & Japan \\ 
72 & CLUMEQ - UniversitÃ© Laval & Canada \\ 
73 & US Army Research Laboratory (ARL) & United States \\ 
74 & CSC (Center for Scientific Computing) & Finland \\ 
75 & DOE/NNSA/LLNL & United States \\ 
76 & KTH - Royal Institute of Technology & Sweden \\ 
77 & University of Minnesota/Supercomputing Institute & United States \\ 
78 & National Centers for Environment Prediction & United States \\ 
79 & National Centers for Environment Prediction & United States \\ 
80 & Rensselaer Polytechnic Institute, Computational Center for Nanotechnology Innovations & United States \\ 
81 & US Army Research Laboratory (ARL) & United States \\ 
82 & NCSA & United States \\ 
83 & University of Southampton & United Kingdom \\ 
84 & NASA/Ames Research Center/NAS & United States \\ 
85 & Clemson University & United States \\ 
86 & NACAD/COPPE/UFRJ & Brazil \\ 
87 & Barcelona Supercomputing Center & Spain \\ 
88 & IBM Poughkeepsie Benchmarking Center & United States \\ 
89 & KTH - Royal Institute of Technology & Sweden \\ 
90 & NCAR (National Center for Atmospheric Research) & United States \\ 
91 & National Institute for Fusion Science (NIFS) & Japan \\ 
92 & Leibniz Rechenzentrum & Germany \\ 
93 & ERDC MSRC & United States \\ 
94 & Indian Institute of Tropical Meteorology & India \\ 
95 & Cray Inc. & United States \\ 
96 & University of Tokyo/Human Genome Center, IMS & Japan \\ 
97 & Cray Inc. & United States \\ 
98 & Georgia Institute of Technology & United States \\ 
99 & NNSA/Sandia National Laboratories & United States \\ 
100 & Commissariat a l'Energie Atomique (CEA) & France \\ 
\end{longtable}
\end{center}

\section{List of scientists who inspired this research agenda}

\subsection{Austria}

\begin{itemize}
\item Alois Ferscha (Johannes Kepler University Linz): ICT systems, pervasive adaptation
\item Stefan Thurner (Vienna): Econophysics, Multi-Player On Line Games
\end{itemize}

\subsection{Cyprus}

\begin{itemize}
\item Anastasia Sofroniou (University of Nicosia): Nonlinear Dynamics, Climate Policies
\end{itemize}

\subsection{Denmark}

\begin{itemize}
\item Kim Sneppen (Niels Bohr Institute): Complex systems, interdisciplinary physics
\end{itemize}

\subsection{France}

\begin{itemize}
\item Henri Berestycki (EHESS, France): Managing Complexity and Institutional Design
\item Guillaume Deffuant (Cemagref): Social Simulation
\item Alan Kirman (Universite Paul Cezanne): Economics
\item Jean-Pierre Nadal (CNRS and EHESS): Complex Systems
\end{itemize}

\subsection{Germany}

\begin{itemize}

\item Thomas Brenner (Philipps-University Marburg): Econophysics
\item Holger Kantz (Max Planck Institute for Physics of Complex Systems,Dresden): Complex Systems, Time Series Analysis
\item J\"{u}rgen Kurths (Potsdam Institute of Climate Research): Interdisciplinary Physics, Sustainability, Climate Policies
\item Thomas Lux (University of Kiel, Institut f\"ur Weltwirtschaft): Economics
\item Carlo J\"{a}ger (Potsdam Institute for Climate Impact Research): Sustainability, Climate Policies
\item J\"urgen Jost (Max Planck Institute for Mathematics in the Sciences): Complex Systems, Networks, Autonomous Systems
\item Kristian Kersting (Fraunhofer): Data Mining
\item Oliver Kirchkamp (University of Jena): Empirical and Experimental Economics
\item J\"{o}rn Kohlhammer (Fraunhofer IGD): Information Visualization
\item Paul Lukowicz (Universit\"{a}t Passau): Sensors networks
\item Kai Nagel (TU Berlin): Large Scale Agents Modeling
\item Albrecht Schmidt (University of Duisburg-Essen): Novel Interfaces for  Ubiquitous Computing
\item Gerhard Tr\"{o}ster (ETH Z\"{u}rich): Sensor Networks, Wearable Computing
\item Peter Wagner (DLR, Germany): Traffic Modeling and Mobility
\item Uwe W\"{o}ssner (Stuttgart): Virtual Reality

\end{itemize}

\subsection{Greece}

\begin{itemize}
\item Panos Agyrakis (University of Thessaloniki): Computational Physics
\item Anastassios Bountis (University of Patras): Complex Systems
\end{itemize}

\subsection{Great Britain}

\begin{itemize}
\item Mike Batty (University College London): Cities, urban development
\item Steven Bishop (University College London): Sustainability
\item Dave Cliff (University of Bristol): Large Scale Complex IT Systems
\item Nigel Gilbert (University of Surrey): Sociology, agent-based modeling
\item Simon G\"{a}chter (University of Nottingham): Behavioural and Experimental Economics
\item Tim Lenton (University of East Anglia): Environmental Modeling
\item Eve Mitleton-Kelly (London School of Economics): Complex Social System
\item Paul Ormerod (British Academy for the Social Sciences/Volterra Consulting): Economics, Networks, Complex Systems
\item Ben Paechter (Edinburgh Napier University): ICT systems, pervasive adaptation, evolutionary computing
\item Jeremy Pitt (Imperial College London): Multi-Agent Systems
\item Tom Snijders (University of Oxford): Multilevel analysis, social statistics, social networks
\item Roger Whitaker (University of Cardiff): Telecommunication Systemss
\end{itemize}

\subsection{Hungary}

\begin{itemize}
\item David Hales (Szeged University): Techno-Social Systems
\item Janos Kertesz (Budapest): Mining of Socio-Economic Data
\item Imre Kondor (Collegium Budapest): Econophysics
\item Andras L\"orincz (E\"otv\"os Lorand University): ICT Systems
\item Gabor Vattay (E\"otv\"os Lorand University): ICT Systems
\end{itemize}

\subsection{Ireland}

\begin{itemize}
\item Petra Ahrweiler (University College Dublin): Innovation Networks
\end{itemize}

\subsection{Italy}

\begin{itemize}
\item Armando Bazzani (University of Bologna): Traffic and transport, GPS data
\item Guido Caldarelli (National Research Council of Italy): Econophysics
\item Anna Carbone (Politecnico di Torino): Econophysics
\item Silvano Cincotti (University of Genoa): Agent-based Economics and Finance
\item Domenico Delli Gatti (University Cattolica): Economics
\item Santo Fortunato (ISI Torino): Information science, sociophysics
\item Mauro Gallegati (University Politecnica delle Marche, Ancona): Economics
\item Vittorio Loreto (University La Sapienza): Language Dynamics, On Line Social Dynamics
\item Rosario Mantegna (Palermo University): Econophysics
\item Matteo Marsili (Abdus Salam ICTP): Econophysics
\item Luciano Pietronero (La Sapienza, Rome): Complex systems, interdisciplinary physics
\item Alex Vespignani (ISI Torino): Techno-Social Systems, Modeling of Epidemic Spreading
\end{itemize}

\subsection{Netherlands}

\begin{itemize}
\item Vincent Buskens (Utrecht University): Economic and Mathematical Sociology, Game Theory
\item Andreas Flache (Univ. Groningen): Computer models of social systems
\item Cars Hommes (University of Amsterdam): Experimental economics
\item Johan Pouwelse (TU Delft): ICT systems, Peer-to-Peer Systems
\item Andrea Scharnhorst (KNAW): Information Systems
\item Bert de Vries (Utrecht University): Sustainable Development Modeling.
 
\end{itemize}

\subsection{Poland}

\begin{itemize}
\item Janusz Holyst (Warsaw Technical University): Sociophysics
\item Andrzej Nowak (Warsaw University): Psychology, complex systems
\end{itemize}

\subsection{Portugal}

\begin{itemize}
\item Jorge Lou\c{c}\~a (Lisbon University): Text-mining, Complex systems.
\item Jos\'{e} Fernando Mendes (University of Aveiro): Interdisciplinary Physics, Complex Networks
\end{itemize}

\subsection{Romania}

\begin{itemize}
\item Carmen Costea (ASE Bucharest): Complex Economic Systems
\end{itemize}

\subsection{Spain}

\begin{itemize}
\item Alex Arenas (Universitat Rovira i Virgili): Econophysics
\item Antonio Ruiz de Elvira (Science and Climate): Climate Policies
\item Maxi San Miguel (IFISC): Sociophysics
\item Anxo Sanchez (Universidad Carlos III de Madrid): Sociophysics
\end{itemize}

\subsection{Sweden}

\begin{itemize}
\item Fredrick Liljeros (Stockholm University): Social Networks, Network Epidemiology, Complexity Theory, Socio Physics
\item Kristian Lindgren (Goteborg University): Complex systems
\item Bj\"orn Ola-Linn\'{e}r (Linköping University): Climate Policies
\item Martin Rosvall (Umea University): Interdisciplinary Physics, Information Science
\end{itemize}

\subsection{Switzerland}

\begin{itemize}
\item Karl Aberer (EPFL): Decentralized ICT systems
\item Kay Axhausen (ETH Zurich): Transportation
\item Wolfgang Breymann (Zurich University of Applied Sciences): Complex Systems
\item Lars-Erik Cederman (ETH Z\"urich): Conflict research
\item Axel Franzen (Bern University): Environmental Sociology
\item Nikolas Geroliminis (EPFL): Traffic systems
\item Dirk Helbing (ETH Zurich): Complex systems, Sociology, Coping with Crises
\item Hans Herrmann (ETH Zurich): Interdisciplinary Physics, Network Theory
\item Thomas Schulthess (ETH Zurich): Supercomputing
\item Frank Schweitzer (ETH Zurich): Econophysics, Network Analysis
\item Didier Sornette (ETH Zurich): Econophysics, Financial Crisis Observatory
\item Yi-Cheng Zhang (University of Fribourg): Econophysics

\end{itemize}

\subsection{U.S.}

\begin{itemize}

\item Xavier Gabaix (New York University): Behavorial economics
\item Sander E. van der Leeuw (Arizona State University): Anthropology
\item Michael Macy (Cornell University): Collective Action, Conflicts, Lab Experiments
\item Paul Werner (California School of Professional Psychology): Personality and Family Assessment, Clinical Decision Making 
\item Douglas White (University of California Irvine): Social Networks, Long-Term Multilevel Study of Human Social Organizations 

\end{itemize}

\printbibliography

\end{document}